\documentstyle[prl,epsfig,aps]{revtex}

\begin{document}

\twocolumn[ \hsize\textwidth\columnwidth\hsize\csname @twocolumnfalse\endcsname

\draft \preprint{\today}

\title{Charge-density wave formation in Sr$_{14}$Cu$_{24}$O$_{41}$}

\author{B. Gorshunov\cite{permanent1}, P. Haas, T. ~R\~o\~om\cite{Taddress}, M. Dressel\cite{email}}
\address{1. Physikalisches Institut, Universit\"at Stuttgart, D-70550 Stuttgart, Germany}
\author{T. Vuletic, B. Hamzic, S. Tomic}
\address{Institute of Physics, Zagreb, Croatia}
\author{J. Akimitsu, T. Nagata\cite{present}}
\address{Department of Physics, Aoyama-Gakuin University, Tokyo, Japan}

\date{Received }
\maketitle

\begin{abstract}
The electrodynamic response of the  spin-ladder compound Sr$_{14-x}$Ca$_x$Cu$_{24}$O$_{41}$ ($x=0, 3, 9$) has been
studied from radiofrequencies  up to the infrared. At temperatures  below 250~K a pronounced absorption peak
appears around $12$~cm$^{-1}$ in Sr$_{14}$Cu$_{24}$O$_{41}$ for the radiation polarized along the chains/ladders
(${\bf E}\parallel {\bf c}$). In addition a strongly temperature dependent dielectric relaxation is observed in
the kHz - MHz range. We explain this behavior by a charge density wave which develops in the ladders sub-system
and produces a mode pinned at $12$~cm$^{-1}$. With increasing Ca doping the mode shifts up in frequency and
eventually disappears for $x=9$ because the dimensionality of the system crosses over from one to two dimensions,
giving way to the superconducting ground state under pressure.
\end{abstract}


]
Low-dimensional quantum spin systems with spins and charges arranged in chains and/or ladders are
under intensive study theoretically as well as experimentally \cite{Dagotto1999}. A number of intriguing magnetic
and electronic properties are inherent to those systems because of their specific arrangement of spins and
charges; also the interplay between spin and charge degrees of freedom produces a variety of unusual phenomena
\cite{Takigawa1998,Ammerahl2000,Kataev2001}. Most important, it was predicted \cite{Rice1993,Dagotto1996} that due
to the presence of a spin gap and paired charge carriers, doped even-leg ladder compounds can produce
superconductivity reminiscent to that in the underdoped high-T$_c$ cuprates; alternatively, they can undergo a
charge-density-wave (CDW) transition.

The spin-ladder system Sr$_{14-x}$Ca$_x$Cu$_{24}$O$_{41}$, where the superconductivity was discovered under
pressure \cite{Uehara1996}, has attracted much attention in the last years. Its structure contains mutually
interpenetrating subsystems of CuO$_2$ chains and Cu$_2$O$_3$ two-leg ladders \cite{Siegrist1988,McCarron1988}.
The parent compound Sr$_{14}$Cu$_{24}$O$_{41}$ is inherently doped with holes and has a rather high dc
conductivity along the $c$ axis (ladder/chain direction) of about $300 ~(\Omega{\rm cm})^{-1}$ at 300 K, with an
anisotropy $\sigma_c$:$\sigma_a$:$\sigma_b\approx$ 1000:100:1 \cite{Motoyama1997} ($a$ denotes the rung direction
and $b$ is the direction perpendicular to the ladders' plane). When Sr is substituted by the isovalent Ca the
conductivity along all three axes increases. It is under debate whether the electronic properties of
Sr$_{14-x}$Ca$_x$Cu$_{24}$O$_{41}$ are determined by a single-particle response or can be influenced by collective
effects, like CDWs \cite{Ruzicka1998,Osafune1997,Kitano2001}. Even in the `simple' Sr$_{14}$Cu$_{24}$O$_{41}$ it
is not clear what the mechanism of dc conductivity is, how the holes redistribute between chains and ladders when
temperature and doping are changed; what is the nature and the interrelation of spin gaps and optical pseudogaps,
the role of collective excitations, and the effect of the dimensionality on charge transport and
superconductivity. In order to address these questions we have investigated the charge dynamics of
Sr$_{14-x}$Ca$_x$Cu$_{24}$O$_{41}$ by means of optical spectroscopy. In this Letter we present clear evidence
that  a CDW forms in this quasi-one dimensional compound. Increasing the dimensionality by Ca doping and pressure
suppresses   the CDW ground state and makes the system superconducting.

The high-quality single crystals (cylinders of about 5~mm in diameter and 7~mm long) were grown by a
traveling-solvent floating-zone method.  For the dc resistivity measurements four electrical contacts to the
sample were obtained with silver paste applied directly on the surface and heated to 600 K in oxygen. In the
radiofrequency range (1 Hz to 1~MHz) the spectra of the real and imaginary parts of the dielectric constant
($\epsilon^{\prime}+i\epsilon^{\prime\prime}$) were obtained from the complex admittance measured at $60~{\rm
K}<T< 110$~K by a technique described in \cite{Silvia}. For frequencies 5 to 25~cm$^{-1}$ a coherent source
quasioptical spectrometer was utilized for direct measurements of $\epsilon^{\prime}$ and
$\epsilon^{\prime\prime}$ \cite{Gorsh}. From the far infrared (FIR) up to 10\,000~cm$^{-1}$ we determined the
polarized reflection by a Fourier transform spectrometer. The combined data sets were analyzed by the
Kramers-Kronig relations in order to obtain the spectra of $\epsilon^{\prime}(\omega)$,
$\epsilon^{\prime\prime}(\omega)$, and conductivity $\sigma(\omega)$. In addition, the non-linear transport of
Sr$_{14}$Cu$_{24}$O$_{41}$ was investigated at various temperatures.

The frequency dependences of the $c$-axis conductivity and the real and imaginary parts of the dielectric
constant of Sr$_{14}$Cu$_{24}$O$_{41}$ are displayed in Fig.~1. Our findings agree with previous infrared
experiments \cite{Ruzicka1998,Osafune1997,Eisaki2000,Osafune1999}. At 300~K the conductivity is Drude-like and
coincides with the dc values. Weak phonon features are seen below 1000~cm$^{-1}$, together with a bump around
2000~cm$^{-1}$ which is caused by the holes in the ladders \cite{Osafune1997,Eisaki2000}. Lowering the
temperature, the overall conductivity from dc up to the FIR decreases making the phonons more pronounced; the
bump looses its intensity. The new features we have observed are the intensive mode in the FIR range at
$10-15~{\rm cm}^{-1}$, and a strongly temperature dependent relaxation in the radiofrequency spectra of
$\epsilon^{\prime}(\omega)$ and $\epsilon^{\prime\prime}(\omega)$. No signs of the FIR mode are found in the $a$
and $b$ directions \cite{Gorsh}. Predictions of a mode existing below 50~cm$^{-1}$ were made earlier on the basis
of infrared measurements in pure \cite{Ruzicka1998} and Ca-doped \cite{Osafune1997} Sr$_{14}$Cu$_{24}$O$_{41}$;
indications of a peak around 14~cm$^{-1}$ were also seen in the room-temperature Raman spectra
\cite{LemmensRaman}. We observe the development of the mode below approximately 250~K. As the temperature
decreases to 60~K (inset of Fig.~1) it gets more pronounced and softens slightly from 14 to 12~cm$^{-1}$;
basically no changes are found going down further to 5~K. While at high temperatures the shape of the mode can be
fitted with a single Lorentzian (plus a Drude term for free carriers), it becomes more complicated at low
temperatures \cite{Gorsh}.

The radiofrequency relaxation is clearly observed as a peak in the $\epsilon^{\prime\prime}$ spectrum accompanied by a
strong dispersion of $\epsilon^{\prime}$. These spectra can be described phenomenologically in terms of a
generalized Debye expressions
\begin{equation}
\epsilon^{\prime}(\omega)+i\epsilon^{\prime\prime}(\omega) = \Delta\epsilon/[1+(i\omega\tau_1)^{1-\alpha}].
\end{equation}
Here $\Delta\epsilon$  is the strength of the relaxation, $\tau_1$ is a mean relaxation time and $1-\alpha$
describes the  relaxation time distribution. The solid lines in Fig.~1 result from fits of the spectra by Eq.~(1).
The so-obtained temperature dependences of the relaxation parameters are shown in Fig.~2 together with the dc
conductivity of Sr$_{14}$Cu$_{24}$O$_{41}$.


We ascribe the 12~cm$^{-1}$ mode to the CDW which develops on the ladders, where the majority of delocalized
charges sit \cite{Osafune1997}, and is pinned by imperfections; correspondingly, we assign the low frequency mode
to the CDW relaxation broadened due to normal carriers, which flow around the pinned CDW and produce screening
currents. Before we come to a detailed analysis, let us summarize the arguments for our interpretation: (a)~ It
is unlikely that the discovered FIR mode is of simple phonon origin. While practically no change is seen in the
phonon spectrum of Sr$_{14-x}$Ca$_x$Cu$_{24}$O$_{41}$ (above 100~cm$^{-1}$) when going from $x=0$ to $x=3$
\cite{Gorsh}, this mode shows a strong shift from 12 to 23 ~cm$^{-1}$ (Fig.1) and disappears with further doping
($x=9$). We also rule out that the resonance is produced by pairs which form a correlated Luttinger liquid -- a
possibility proposed recently \cite{Kim2001} -- because the mode appears at elevated temperatures and cannot be
fitted by the corresponding expressions. (b)~The CDW is one of the  possible ground states in a spin-ladder
system like Sr$_{14}$Cu$_{24}$O$_{41}$ \cite{Dagotto1999,Dagotto1996} where the electronic bands are essentially
of one-dimensional character~\cite{Arai1997}. (c)~We observe the CDW-related absorptions only along the highly
conducting $c$ axis. (d)~As expected for the formation of a CDW \cite{Gruner}, experimental indications for the
charge order \cite{Eisaki2000} and lattice superstructure \cite{Takahashi1997} are seen in the ladders of
Sr$_{14-x}$Ca$_x$Cu$_{24}$O$_{41}$. The phonon spectrum shows signatures of a Brillouin zone folding below 200~K
\cite{Gorsh} which gives evidence for a lattice superstructure. (e)~The resonance at 12~cm$^{-1}$ is characterized
by an enhanced effective mass, typical for collective excitations like the CDW. (f)~We do not observe any
resonance for $x=9$ when the system looses its one-dimensionality \cite{Nuker2000,Isobe}. (g)~The absorption
features we observe in Sr$_{14}$Cu$_{24}$O$_{41}$ are extremely reminiscent of the CDW response in most
well-known one-dimensional compounds like K$_{0.3}$MoO$_3$, NbSe$_3$, TaS$_3$, and (TaSe$_4$)$_2$I \cite{Gruner}.
As we will discuss next, this holds qualitatively (Fig.~1) as well as quantitatively.

Utilizing a standard theory for one-dimensional CDW systems we now will describe the dynamical response of
Sr$_{14}$Cu$_{24}$O$_{41}$. Following Littlewood's model \cite{Littlewood} we consider a deformable CDW which
interacts with the uncondensed carriers in conditions of non-uniform pinning. It accounts for the existence of two
modes: transverse and longitudinal. The former couples to electromagnetic radiation and yields a high frequency
pinned CDW mode, whereas the latter couples to an electrostatic potential and results in an overdamped low
frequency relaxation around $1/(2\pi \tau_1)$ due to screening effects. In the real systems, absorptions arise
from  the mixing of a longitudinal plasma-like oscillations of the CDW (with the screened plasma frequency
$\Omega _p^{2}=\rho_c^{2}/\epsilon_zm^{*}$) with a transverse pinned  mode at $\Omega_0=\sqrt{V_0/m^*}$ due to
symmetry breaking (non-uniform pinning); this mixing can account for the unusual spectral shape of the CDW
resonance we observe in Sr$_{14}$Cu$_{24}$O$_{41}$. The pinning potential is given by $V_0$; the charge density
by $\rho_c=ne$; $\epsilon_z$ is the high-frequency dielectric constant, and $m^*$ is the effective CDW mass.
First we note that the position of the relaxation peak in $\epsilon^{\prime\prime}(\omega)$ is thermally
activated (Fig.~2) with the same activation energy as the dc conductivity, as predicted by the model
\cite{Littlewood}. Further, estimating $n\approx 10^{21}~{\rm cm}^{-3}$  as the carrier concentration
\cite{Osafune1999}, taking $\sigma_z =10^{-4}~(\Omega{\rm cm})^{-1}$ as the plateau conductivity between the two
excitations and $1/(2\pi c\tau_1 )=1.7\times 10^{-5}~{\rm cm}^{-1}$, $\Omega_0/(2\pi c) =12~{\rm cm}^{-1}$
(Fig.~1), $\epsilon_z=5$ \cite{Gorsh}, we get an effective mass $m^*\approx 200m_0$, a plasma frequency
$\Omega_p/(2 \pi c) \approx 190~{\rm cm}^{-1}$, and $n^{*}=m^{*}\Omega_{p}^{2}/(4\pi e^{2}) \approx 2\times
10^{19}~{\rm cm}^{-3}$ carriers condensed, i.e., only 2~\%\ of the charge carriers contribute to the CDW. The CDW
plasma frequency can be independently estimated from the oscillator strength of the mode $\Omega_p/(2\pi c)
=\Omega_0/(2\pi c\sqrt{\epsilon_z})=120~{\rm cm}^{-1}$ \cite{remark1}, which is in good agreement with the value
of $190~{\rm cm}^{-1}$ obtained above.

We assign the low-temperature decrease of the spectral weight in the FIR (below 1000~cm$^{-1}$, see Fig.~1 and
\cite{Eisaki2000}) to the opening of a CDW pseudogap $E_g/hc\approx 2000~{\rm cm}^{-1}$ (i.e.,\ $E_g\approx
0.27$~eV) and determine the CDW mean-field temperature $T_{\rm CDW}^{MF}=E_g/3.5k_B\approx 900$~K; this implies
that CDW fluctuations should be observed well above the room temperature. Similar rather high mean-field
temperatures were estimated in other quasi one-dimensional conductors like TTF-TCNQ  ($T_{\rm CDW}^{MF}\approx
450$~K)
or K$_{0.3}$MoO$_3$ (330~K) \cite{Gruner}. The pseudogap should govern the activated behavior of the dc
conductivity (Fig.~2, $170~{\rm K}<T<250$~K). The fact that $\sigma_{dc}(T)$ is not simply activated for $250~{\rm
K}<T<300$~K can be explained with CDW fluctuations which produce a collective contribution to the dc transport.
This sort of paraconductivity was also detected in other one-dimensional CDW conductors, like K$_{0.3}$MoO$_3$ or
TTF-TCNQ \cite{Gorshunov1994}.

Below approximately 170~K the activation energy of $\sigma_{dc}(T)$ changes from $E_g=0.27$~eV to 0.12~eV. A
number of magnetic, electronic and structural properties display an anomaly around 170~K
\cite{Kataev2001,Isobe,Zhai,Cox1998,Owen2000} which may indicate some type of phase transition. We suggest that
this transition is due to a second CDW which develops in the chain sub-system of Sr$_{14}$Cu$_{24}$O$_{41}$.
Mobile holes in the chains, essential for the CDW formation, have been suggested on the basis of ESR measurements
\cite{Kataev2001}. The additional peak observed in the microwave range at $1.7~{\rm cm}^{-1}$ for $T<170$~K
\cite{Kitano2001,Maeda}, shown in Fig.~1, thus corresponds to the second pinned CDW mode; its spectral weight is
much smaller compared to the FIR mode, due to the lower charge concentration in the chains. Similar to the FIR
mode reported here, this microwave resonance is seen in pure Sr$_{14}$Cu$_{24}$O$_{41}$ and for small Ca doping
of $x=3$ while is disappears for larger $x$. We note that signs of this resonance lying below 8~cm$^{-1}$ are
also clearly seen in our low temperature reflectivity spectra \cite{Gorsh} for $x=3$.

Although the presented results clearly point towards the collective CDW ground state in the spin-ladder system
Sr$_{14}$Cu$_{24}$O$_{41}$ and can even be well described by Littlewood's model, a few open questions remain. A
signature of an absorption line at 14~cm$^{-1}$ was already observed in the Raman spectra at room temperature
\cite{LemmensRaman}. While the CDW phase fluctuations should be infrared active and the CDW amplitude
fluctuations  only Raman active \cite{Gruner}, it is likely that both excitations mix  due to symmetry breaking;
hence the resultant excitation is observable by both techniques (but still only along the conducting $c$
direction). Interestingly, the Raman mode seems to be pinned  even at 300~K, while we do not see any sign of the
optical resonance at this temperature probably due to the large screening by free carriers. The CDW can be
depinned by an external electric field leading to non-linear transport \cite{Gruner}. We do observe a
field-dependent conductivity in our system, however, the non-linearity is very small and the threshold current
cannot be clearly determined. An estimate of the threshold field $E_T\approx0.1$~V/cm (Fig.~2) cannot be related
to the pinning frequency $\Omega_0$  by the standard expression \cite{Gruner} $m^{*}\Omega^{2}_0=2eE_T/\lambda$,
with the wavelength $\lambda$ of the CDW taken of the order of the distance between Cu atoms on the ladder
($\approx$4~\AA); a similar discrepancy was noticed for the microwave mode \cite{Kitano2001}. The effective mass
$m^{*}\approx200m_0$ is small compared to standard CDW materials where up to the order of ten thousand has been
reported \cite{Gruner}. The mentioned points can be ascribed to an unusual CDW state formed in
Sr$_{14}$Cu$_{24}$O$_{41}$, as compared to typical one-dimensional conductors. This is most likely due to the two
interacting sublattices on the chains and on the ladders. Since each is characterized by magnetic, structural,
and charge ordering, the ground state of the system is determined by a mixture of correspondent order parameters
leading to excitations which may be rather unconventional. Detailed low-energy studies of the interplay between
the spin and charge subsystems in (Sr,Ca,La)$_{14}$Cu$_{24}$O$_{41}$ for different  doping and in magnetic fields
are required. In addition, the internal deformations of the CDW and the pinning mechanism may be more complicated
in this system of interacting chains and ladders compared to strictly one-dimensional compounds. It is not
surprising that the observed CDW phenomenon is strongly modified compared to the conventional CDW compounds.


In summary, we presented evidence  of a charge-density wave formation in the ladder subsystem in
Sr$_{14-x}$Ca$_x$Cu$_{24}$O$_{41}$. For $x=0$ we observed a pinned CDW mode at 12~cm$^{-1}$ below 250 K and a
strongly temperature dependent relaxation in the radiofrequency range due to single-particle screening currents.
The pinned mode shifts up to 23~cm$^{-1}$ for $x=3$ and disappears when the Ca doping reaches $x=9$. From the two
competing ground states (density wave and superconductivity) expected for spin-ladder compounds, the CDW
instability dominates when Sr$_{14-x}$Ca$_x$Cu$_{24}$O$_{41}$ is highly one-dimensional ($0\leq x\lesssim 3$) but
shows a tendency to disappear when stronger Ca doping ($x=9$) makes it more two-dimensional. When external
pressure increases the coupling (and thus the dimensionality) even further, the system finally becomes
superconducting.

We thank P. Lemmens and A. Maeda for discussions and exchange of results prior to publication.
The work was supported by the Deutsche Forschungsgemeinschaft (DFG)
and Ministry of Science of RH.


Fig.1. (a) Frequency dependent conductivity of Sr$_{14}$Cu$_{24}$O$_{41}$ measured at different temperatures
along the highly conducting $c$ direction. The arrows indicate the dc conductivity; the solid lines represent
fits by the generalized Debye relaxation, the dashed lines are guides to the eye. The peak around 1~cm$^{-1}$
corresponds to the data from \cite{Kitano2001}. The inset shows how the shape of the pinned CDW resonance varies
with temperature.
 (b)~Real and imaginary parts of the dielectric constant as a function of frequency;
the solid lines are fits by the generalized Debye expression.
The inset exhibits the  evolution of the spectral weight in the FIR range with decreasing temperature:
the dotted line shows the pinned CDW mode of
Sr$_{11}$Ca$_3$Cu$_{24}$O$_{41}$.

Fig.2. Temperature dependences of the dc conductivity of
Sr$_{14}$Cu$_{24}$O$_{41}$ along the $c$ axis
and of the parameters which characterize the dielectric relaxation at radiofrequencies [Eq.~(1)]:
the inverse time $1/(2\pi \tau_1)$ and
the strength $\Delta \epsilon$
of the dielectric relaxation, and the parameter 1-$\alpha$ which describes the
symmetrical broadening of the relaxation time distribution.
The dc conductivity is thermally activated
$\sigma_{dc}\propto {\rm exp}\{-\Delta/k_BT\}$ between 250~K and 170~K with
an energy gap $\Delta\approx 0.27$~eV  ($\Delta/hc\approx 2200$~cm$^{-1}$) in agreement with the onset of  the
low-temperature conductivity spectrum in  Fig.~1 (inset of the lower panel)
and is identified as a pseudogap due to CDW in the ladders.
Below 170 K a change of the activation energy to $\Delta\approx 0.12$~eV
is observed due to the onset of a CDW in the chains.
The solid lines correspond to an activated behaviors of $\sigma_{dc}(T)$ and $1/\tau_1(T)$.
The inset shows an electric field dependent  resistance of
Sr$_{14}$Cu$_{24}$O$_{41}$ measured along the $c$ direction at 87~K.


\begin{thebibliography}{99}
\bibitem{remark1}
We use the dielectric contribution of the mode $\Delta\epsilon \approx 100$
estimated from the directly measured $\epsilon^{\prime}$ and $\sigma$ spectra,
shown in the inset of Fig.~1.
\end{thebibliography}
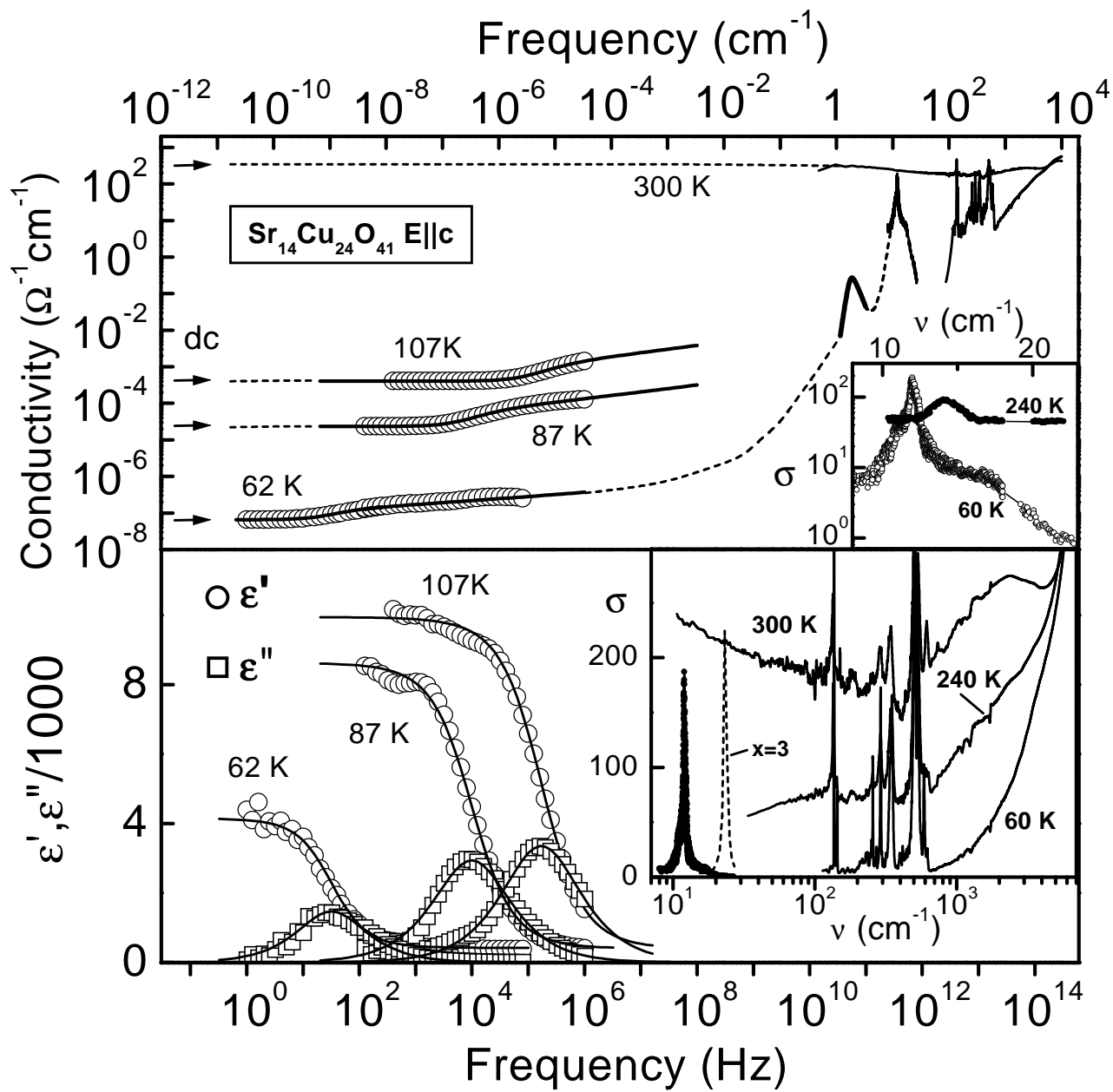
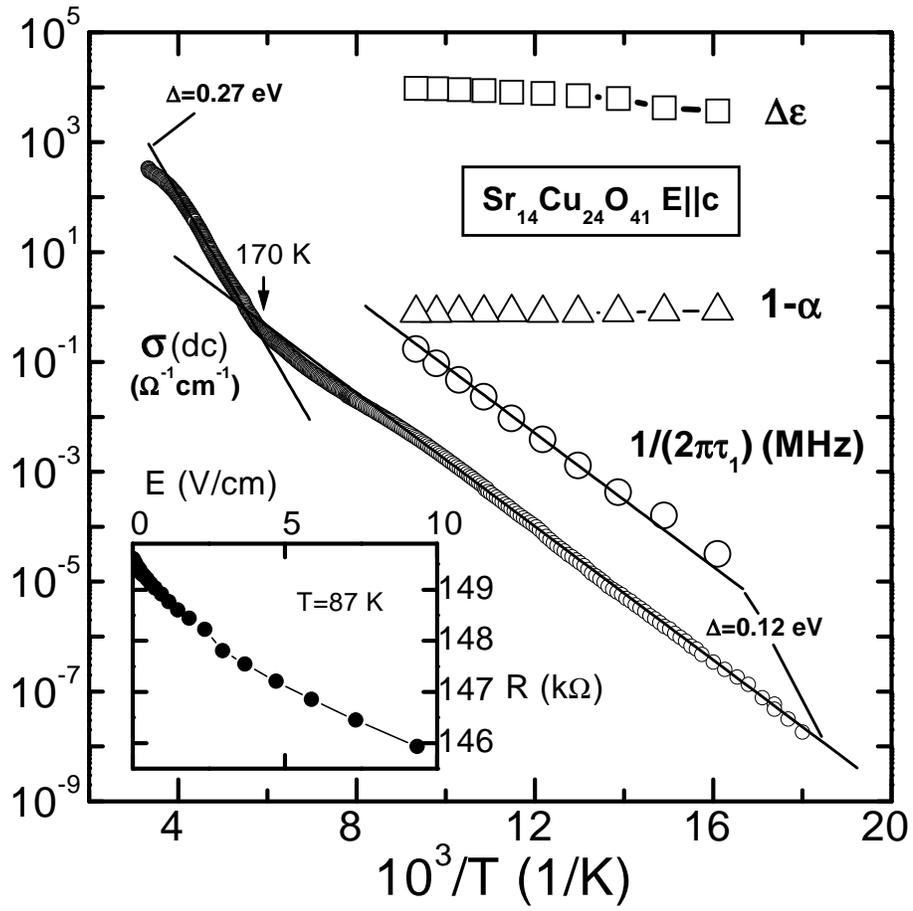
\end{document}